\documentclass[aps,prl,preprint,groupedaddress,showpacs]{revtex4}
\usepackage{graphicx}
\usepackage{amssymb}
\begin{document}
\title{VOLUME REFLECTION AND REFRACTION OF RELATIVISTIC PARTICLES IN BENT CRYSTALS}

\author{Gennady V. Kovalev}

\affiliation{School of Mathematics \\University of Minnesota, Minneapolis, MN 55455,USA}

\begin{abstract}
The quasi-channeling of positive and negative relativistic particles in a bent crystal is studied using the classical deflection function.  It was shown that the potential scattering in a central field of  bounded  ring-like potentials may produce the ``reflected'' and ``refracted'' fractions of scattered particles. For particles with positive charge the ``reflected'' fraction is mainly presented; at the same time we predict that for particles with negative charge the ``refracted'' fraction should dominate. The effect of ``empty core'' for central scattering is also discussed. The average deflection angles for volume ``reflection'' and ``refraction'' are derived for accepted potential model of the crystal. The calculated average ``reflection'' angle  is in satisfactory agreement  with  recent experimental data \cite{ivanov_2006}. 
\end{abstract}



\maketitle
\section{\label{sec:01} Introduction}
In the early 80-s, the experiments \cite{andreev_1982,andreev_1984,andreev_1986} have discovered the effect of volume capture of relativistic particles in a channeling motion by uniformly bent crystal. Several attempts have been made to find a specific mechanism of capture (see e.g.\cite{tar98}) because the  coherent  collisions in a central field can not cause the channeling. Taratin and Vorobiev studied the passage of high energy particles in a bent crystal using the computer simulations  \cite{tar87_1,tar87_2} and found that the particles (both positive and negative) are deflected to the convex side the bend of the atomic planes by an angle of about $2*\theta_c$ ($\theta_c$ - critical angle). They called this a volume ``reflection'' by bent atomic planes.  The significant beam divergence and multiple scattering of the quasi-channeled particles in the crystals prevented observing  this effect for almost two decades. The experiment \cite{ivanov_2005,ivanov_2006} of $70 GeV$ protons scattered by a short bent crystal is the first explicit demonstration of the volume ``reflection''. It should also be mentioned that preliminary results of crystal collimation at RHIC \cite{fliller_2005} for the 2001 run demonstrated some interesting results which were recently explained \cite{bir_2006_1,bir_2006_3} by the same effect. 

My purpose here is to work out the analytic formulas for volume ``reflection'' and related effects, and discuss the principle difference between deflections of positive and negative particles. The standard continuous screening potential for the bent crystal is assumed to be a good approximation. The small angle formula for classical scattering \cite{ll1} are not applicable here, because the continuous potential could not be considered weak near the turning points although the angles of deflection are small. This article will deal with exact solution of potential scattering by specific model of the crystal's potential.    

\section{\label{sec:02}  Models of Potential for Bent Quasi-Channeling and Channeling}
The charged particles incident on a bent single crystal along the tangent directions to the crystallographic planes, experience a collective field from atoms of the tangent areas of these  planes. So, the  starting point for studying the quasi-channeling and channeling in bent crystals is the interaction potential for binary ion-atom collisions, 

\begin{eqnarray}
U_{atom}(r)=\frac{Z_P Z_T}{r}f(\frac{r}{R_{T-F}}),
\label{atom_potent}
\end{eqnarray}
where $Z_{P,T}$ are the bare nuclear charge of the projectile and crystal atoms, $R_{T-F}$ is  Thomas-Fermi (or Firsov) screening length and $f$ is a screening function,  
\begin{eqnarray}
f(\frac{r}{R_{T-F}})=\sum_{i} \alpha_i exp(\beta_i\frac{ r}{R_{T-F}}).
\label{screening_factor}
\end{eqnarray}
The screening function approaches $1$ in the limit $r \rightarrow 0$ and decays 
exponentially for large $r$. This is only  possible if $\sum_{i} \alpha_i=1$. 
Different analytic forms of the screening function $f$ are in use. The 
Moliere approximation\cite{moliere_1947}, $f_M$, uses $\{\alpha_i\} = \{0.1,0.55,0.35\}$,   $\{\beta_i\} = \{6.0,1.2,0.3\}$. Other choices include the Ziegler-Biersack-Littmark 
potential\cite{zbl_1985} $f_{ZBL}$ with $\{\alpha_i\} = \{0.1818, 0.5099, 0.2802, 0.0281\}$, and $\{\beta_i\} = \{3.2, 0.9423, 0.4028, 0.2016\}$. Now, if we find the average potential over a cylinder with the radius $R$ and the root-mean-square temperature  displacement $u$ of atoms in gaussian form, we receive the following expression for continuous cylindrical potential  
\begin{eqnarray}
U_{p}(\rho)=
\overline{U}\sum_{i}\frac{\alpha_i}{\beta_i}\exp(\beta_i^2\frac{u^2}{2R_{T-F}^2})
(\exp(\beta_i\frac{|\rho-R|}{R_{T-F}})erfc(\beta_i\frac{u}{\sqrt{2}R_{T-F}}+\nonumber\\ \frac{|\rho-R|}{\sqrt{2}u})
+\exp(-\beta_i\frac{|\rho-R|}{R_{T-F}})erfc(\beta_i\frac{u}{\sqrt{2}R_{T-F}}-\frac{|\rho-R|}{\sqrt{2}u})).
\label{bent_PP}
\end{eqnarray}
Here $\rho$ is polar coordinate from center of cylinder, $\overline{U}=\pi Z_P Z_T R_{T-F} n_p$, $n_p$ is a density of atoms in the bent plane, $erfc()$ is the complementary error function\cite{abst_1964}. 

For charged projectiles , the screen potential of atoms (\ref{atom_potent}) over all space is positive or negative depending on the sign of $Z_p$. It follows that the average potential of each plane $U_{p}(\rho)$ and the whole cylindrical crystal potential $U(\rho)=\sum U_{p}(\rho)$ will also be positive or negative depending on the charge of the projectile (Fig.1 (A)). In the intervals $0< \rho < R-Nd$,  $R< \rho < \infty$ the potential $U(\rho)$ decreases to zero. In the interval $R -Nd < \rho < R $, potential does not have a singularity and is smooth almost periodic function with some constant positive or negative pedestal. The value of pedestal is quite significant. It is  ~13.3 eV for Si $<110>$ plane (T=300K) with maximum peak of the potential ~35.9 eV, and ~10 eV for Si $<111>$  with maximum peak ~34.7 eV, i.e. about 1/3  of maximal value of the screening potential. 

When the particles enter the bent crystal from outer side, the influence of the pedestal might be significant because the outer potential ring has bigger magnitude.
Most deflection experiments \cite{ivanov_2005,ivanov_2006,fliller_2005} are done using a different geometry when the particles enter the piece of crystal from the front end and the influence of the pedestal in this case is quite small. The value of pedestal can be subtracted and 'effective' potential and its rectangular model are shown in Figs. 1(C),(D).
However, the signs of scattering potentials still remain the same. To estimate the trajectories of particles and  deflection function we will use the following parameters for the crystal Si $<111>$ used in\cite{ivanov_2006}: $U_{o}=25 eV$, $a=0.78 \AA$, $d=3.13 \AA$ (see Fig.1(D)).

\section{\label{sec:03} Classical Deflection Function on Ring Potential}

The path of a particle in a central classical field is, as well known, symmetrical about a line from the center to the nearest point in the orbit (OA for positive particles in Fig.2(A)). 
Hence the two asymptotes to the orbit make equal angles $\varphi$ with these lines. The angle $\chi=\pi-2\varphi$ through which the particles are deflected as they pass the center of the potential varies in the interval $-\pi < \chi< \pi$ when the angle $\varphi$ varies in the interval $0 < \varphi< \pi$.  For relativistic particles, the angles $\varphi$ are near $\pi/2$ and $\chi \approx 0$,  so it is convenient to introduce the angle $\alpha =\pi/2 - \varphi$ (see Fig.2 (A), (B), (C)) which is positive when the particle deflected to the left in Fig.2(A) and negative when the particle deflected to the right.  At high energies, there is no ``orbiting'' or ``spiral'' scattering \cite{MillerWH_1969_1} where the angle $\varphi$ can have a larger interval.  However, there is always a possibility of different signs for deflection angle $\chi=2\alpha$ in the case of a single ring or periodic system of rings concentrated near a circle, as we will discuss below. 

We shall first consider the deflection of a particle of mass $m$ and energy $E$ passing the boundaries of a positive rectangular ring $U_o >0$  with the outer radius $R$ and thickness $a$ (see Fig.2 (B)). The path of a particle is refracted by the angle $(\theta_0 - \theta_1)$  and $(\beta_1-\beta_0)$  to the incident directions,  at the point of entrance and exit of the ring potential. These angles define  the deflection angle  $\alpha=\chi/2$ for over-barrier (quasi-channeling) motion of the particles:
\begin{eqnarray}	
\alpha(b,E) = (\theta_0-\theta_1)-(\beta_1-\beta_0),
\label{HalfDeflection}
\end{eqnarray}
with the nearest point  $A$ on the path located inside the ring.  
The tangent continuity of a particle's momentum gives the relativistic relations between pair of angles  $\theta_0$, $\theta_1$  and   $\beta_0$, $\beta_1$:

\begin{eqnarray}	
(1-\phi)sin^2(\theta_1) = sin^2(\theta_0)-\phi,\;\label{Refraction1}\\
(1-\phi)sin^2(\beta_1) = sin^2(\beta_0)-\phi,
\label{Refraction2}
\end{eqnarray}

where the 
\begin{eqnarray}	
\phi=\frac{2E U_0}{E^2-m^2c^4}
\label{LindhardAngle}
\end{eqnarray}

is the squared Lindhard's angle $\phi=\pm\theta_c^2$ with the sign of the potential. In deriving (\ref{Refraction1}),(\ref{Refraction2}), we made one assumption, $2EU_0>> U_0^2$, which is, in general, satisfied for nonsingular potentials and wide range of energies including the nonrelativistic. From geometry of scattering (Fig.2 (B)) and relations (\ref{Refraction1}),(\ref{Refraction1}), we can
receive the explicit expressions for angles: 
\begin{eqnarray}	
\theta_0 = arcsin(\sqrt{1-\frac{b^2}{R^2})},\;  \theta_1=arcsin(\sqrt{1-\frac{b^2}{(1-\phi)R^2}} )\\
\beta_0 = arcsin(\sqrt{1-\frac{b^2}{(R-a)^2})},\;  \beta_1=arcsin(\sqrt{1-\frac{b^2}{(1-\phi)(R-a)^2}} ),
\label{AnglesRefraction}
\end{eqnarray}
as functions of impact parameter $b$, energy of particle $E$, and parameters of potential $R$, $a$, $U_0$. 
From now on, we will use the normalized parameters:	
\begin{eqnarray}	
\hat{b} = b/R, \;\;  \hat{a}=a/R,\;\;  \hat{d}=d/R,\;\;  \hat{b}_{a}=b/(R-a)=\hat{b}/(1-\hat{a}).
\label{Normalize1}
\end{eqnarray}
The deflection angle (\ref{HalfDeflection}) then can be written:
\begin{eqnarray}	
\alpha(b,E) =arcsin(\frac{\hat{b}(\sqrt{1-\hat{b}^2}-\sqrt{1-\phi-\hat{b}^2})}{{\sqrt{1-\phi}}})- arcsin(\frac{\hat{b}_{a}(\sqrt{1-\hat{b}_{a}^2}-\sqrt{1-\phi-\hat{b}_{a}^2})}{{\sqrt{1-\phi}}}),
\label{HalfDeflectionGen}
\end{eqnarray}
where the first term correspond to $(\theta_0-\theta_1)$, and second to  $(\beta_1-\beta_0)$ in Eq.(\ref{HalfDeflection}). 

It is worth to note that the Eq. (\ref{HalfDeflectionGen}) was derived for the particles transmitted inside the ring, but it is, in general, valid for other cases as well. In the following, we consider the modification of this expression. 

$Case\; 2P$. If the impact parameter $b$ satisfy the inequality $\sqrt{1-\phi}(1-\hat{a})<\hat{b}<(1-\hat{a})$,   the last radical in second term of (\ref{HalfDeflectionGen})  vanishes. The deflection function  
(\ref{HalfDeflectionGen}) in this interval will be
\begin{eqnarray}	
\alpha(b,E) =arcsin(\frac{\hat{b}(\sqrt{1-\hat{b}^2}-\sqrt{1-\phi-\hat{b}^2})}{{\sqrt{1-\phi}}})-arcsin(\frac{\hat{b}_{a}(\sqrt{1-\hat{b}_{a}^2})}{{\sqrt{1-\phi}}}).
\label{HalfDeflectionCase2P}
\end{eqnarray}
It monotonically increases from negative values to positive (Fig.3 (A)).
  
$Case\; 3P$. If the impact parameter $(1-\hat{a})<\hat{b}<\sqrt{1-\phi}$,  the particle does not pass across the inner boundary of ring and the whole second term in (\ref{HalfDeflectionGen})  vanishes. The deflection function  
(\ref{HalfDeflectionGen}) in this case continues to increase (Fig.3 (A)),
\begin{eqnarray}	
\alpha(b,E) =arcsin(\frac{\hat{b}(\sqrt{1-\hat{b}^2}-\sqrt{1-\phi-\hat{b}^2})}{{\sqrt{1-\phi}}}),
\label{HalfDeflectionCase3P}
\end{eqnarray}
becomes the well known formula for the scattering by a spherical potential 'well' or cylindrical disk.  

$Case\;4P$. For a narrow interval of impact parameters $\sqrt{1-\phi}<\hat{b}<1$ or  $R-b< \frac{R\phi}{2}$, the particles are reflected from the outer boundary of potential and this effect is responsible for the volume ``reflection''. The second radical in (\ref{HalfDeflectionGen})  vanishes in this case and deflection function is given by
\begin{eqnarray}	
\alpha(b,E) =arcsin(\frac{\hat{b}(\sqrt{1-\hat{b}^2})}{{\sqrt{1-\phi}}}).
\label{HalfDeflectionCase4P}
\end{eqnarray}  

In this same manner we can study the trajectories of particles scattered by a negative rectangular ring $U_o < 0$ (Fig.2 (C)), and it gives the exactly same expression (\ref{HalfDeflectionGen}) for deflection function.  The only difference, that $\phi$ is negative, produces just three cases in comparison with positive ring.  

$Case\;2N$.
For negative potential $\sqrt{1-\phi}>1$ and the impact parameter in the interval $(1-\hat{a})<\hat{b}<(1-\hat{a}) \sqrt{1-\phi}$, the first radical of second term in (\ref{HalfDeflectionGen}) becomes zero. The modified deflection function,  
\begin{eqnarray}	
\alpha(b,E) =arcsin(\frac{\hat{b}(\sqrt{1-\hat{b}^2}-\sqrt{1-\phi-\hat{b}^2})}{{\sqrt{1-\phi}}})+arcsin(\frac{\hat{b}_{a}(\sqrt{1-\phi-\hat{b}_{a}^2})}{{\sqrt{1-\phi}}}),
\label{HalfDeflectionCase2N}
\end{eqnarray}
monotonically decreases from positive values to negative (Fig.3 (C)). This interval of impact parameters corresponds to the reflection from inner wall of the negative potential. 

$Case\;3N$. If the impact parameter lies in the interval $(1-\hat{a}) \sqrt{1-\phi}<\hat{b}<1$, the whole second term of (\ref{HalfDeflectionGen}) becomes zero. The particle freely moves inside the negative ring and the form function in this case has the expression (\ref{HalfDeflectionCase3P}) of $Case \; 3P$ . It was pointed out  already that  this gives the formula for classical scattering by a spherical potential well \cite{ll1}. This deflection function is shown in Fig.3(C) by the doted line.
													
The sequence of considered cases can be violated if the radius of the ring $R > R_{c}= \frac{2a}{\|\phi\|}$. Indeed, in the $Case\;3P$, where $\sqrt{1-\phi}<1$, the interval $(1-\hat{a})<\hat{b}<\sqrt{1-\phi}$ vanishes if $R$ increases and $\hat{a}\rightarrow 0$  for positive potentials. In the $Case\;3N$ where $\sqrt{1-\phi}>1$ the interval $(1-\hat{a}) \sqrt{1-\phi}<\hat{b}<1$ vanishes if $R$ increases and $\hat{a}\rightarrow 0$  for negative potentials. In both cases, the interval of transmission the particle without the the touching the inner wall disappearres, and the deflection functions have discontinuances (Fig.3(B),(C)).  This situation takes place in  experiments \cite{ivanov_2005,ivanov_2006}, where  $E=70GeV$, $\phi\cong0.58*10^{-9}$ (Si $<111>$), $2a\cong1.56*10^{-10}m$, and $R_{c}=28cm$. In periodic structure, which we consider below, the potential thickness, $a$, should be substituted by the period of structure, $d$. This gives $R_{c}= \frac{2d}{\|\phi\|}\cong 1m$ which is also less than radius of the bend  $R=1.7m$.  

\section{\label{sec:04} Effect of ``empty Core''}

There is interested feature of the classic scattering in central fields which we would like to discuss here. If the particles are scattering by a positive potential disk (i.e. a field with $U = 0$ for $\rho > R$ and $U = U_o$ for $\rho < R$), the  deflection functions (see Eq.(\ref{HalfDeflectionCase3P})) are strictly positive (doted line in Fig.3(A)) and monotonic in the interval $0<\hat{b}<\sqrt{1-\phi}$, i.e. the ``disk'' scatters strictly as a repulsive potential in this region of impact parameters. We have the opposite situation for scattering by negative potential ``well'' (i.e. a field with $U = 0$ for $\rho > R$ and $U = -U_o$ for $\rho < R$), where all scattering angles are negative (doted line in Fig.3(C)). The ``well'' scatters strictly as a attractive potential. Now we cut off the central circular core of these potentials with radius $R-a$ to receive positive or negative ``rings''. In the positive central ring, all the particles with impact parameters $0<\hat{b}<\sqrt{1-\phi}(1-\hat{a})$ are deflected with a negative sign (Fig.3(A)), i.e. empty core for positive ring scatters as an attractive potential for widest range of impact parameters.  In the negative central ring, all the particles with impact parameters $0<\hat{b}<(1-\hat{a})$ are deflected with positive sign (Fig.3(A)), i.e. empty core here scatters as a repulsive potential. 

The physics of this effect is simple. The inner wall of the positive ring creates a force toward the center of the ring and particles obtain the negative scattering angle. The inner wall of the negative ring creates a force in upward direction and the scattering angle for penetrated particles obtain the positive sign.
However, to some extent the behavior of the particles in the area of  ``empty core'' is independent of the potential form, i.e. different forms of the potential can produce the same effect.  We can observe that a projectile with some impact parameter $b$ and fixed point of exit (point E in Fig.2(B,C)) has already defined direction of motion, no matter what  shape of the potential constitutes the ring.  Indeed, the momentum $p$ and angular momentum $p b$ of projectile must conserve. These imply that the path of the projectile must come out from point of exit $E$ and be tangent to the circle of radius $b$ in the area of  ``empty core'' (Fig. 2(B),(C)). Hence, the point of exit and the impact parameter  enables us to obtain the scattering angle of particle without knowing the the exact shape of potential inside the ring. There is another useful feature of this effect. It produces the coherent action in wide interval of impact parameters greater than the period of ring, so it might be useful for applications.

\section{\label{sec:05} Reflection and Refraction by System of Rings}

Finally lets consider the scattering by the periodic system of the potential rings mentioned in Sec.\ref{sec:02}. The relativistic particle moves with impart parameter $b$  consequently 'refracted' on entering and leaving each potential ring. It is easy to  understand, that the deflection function for this case will be the  sum of the deflection functions of the individual rings (\ref{HalfDeflectionGen}):
\begin{eqnarray}	
\alpha(b,E) = \sum^{N}_{i}\alpha^{(i)}(b,E).
\label{HalfDeflectionS}
\end{eqnarray}
Here
\begin{eqnarray}	
\alpha^{(i)}(b,E) =arcsin(\frac{\hat{b}_{i}(\sqrt{1-\hat{b}^2_{i}}-\sqrt{1-\phi-\hat{b}^2_{i}})}{{\sqrt{1-\phi}}})-\nonumber\\ arcsin(\frac{\hat{b}_{ia}(\sqrt{1-\hat{b}_{ia}^2}-\sqrt{1-\phi-\hat{b}_{ia}^2})}{{\sqrt{1-\phi}}}),
\label{HalfDeflectionGen_i}
\end{eqnarray}
where the normalized impact parameters are

\begin{eqnarray}	
\hat{b}_{i} =\frac{\hat{b}}{1-i \hat{d}} , \; \; \hat{b}_{ia}=\frac{\hat{b}}{1-\hat{a}-i \hat{d}}.
\label{Normalize2}
\end{eqnarray}

In the relativistic case all pairs  $(\sqrt{1-\hat{b}^2_{i}}-\sqrt{1-\phi-\hat{b}^2_{i}})$,    $(\sqrt{1-\hat{b}_{ia}^2}-\sqrt{1-\phi-\hat{b}_{ia}^2})$ are extremely small and substitution $arcsin(x)\rightarrow x$ can be done in (\ref{HalfDeflectionGen})-(\ref{HalfDeflectionGen_i}) without losing the accuracy, e.g. the Eq.(\ref{HalfDeflectionS}) can be rewritten

\begin{eqnarray}	
\alpha(b,E) =\sum^{N}_{i}\frac{\hat{b}_{i}(\sqrt{1-\hat{b}^2_{i}}-\sqrt{1-\phi-\hat{b}^2_{i}})}{{\sqrt{1-\phi}}}- \frac{\hat{b}_{ia}(\sqrt{1-\hat{b}_{ia}^2}-\sqrt{1-\phi-\hat{b}_{ia}^2})}{{\sqrt{1-\phi}}}.
\label{HalfDeflectionS_M}
\end{eqnarray}

If the impact parameter $\hat{b}  \tilde{<} (1-N\hat{d})$, all terms in Eq.(\ref{HalfDeflectionS}), (\ref{HalfDeflectionS_M}) are present and the particles transmit into the empty core of potential (quasi-channeling). The deflection angles in this case were discussed in Sec.\ref{sec:04}. If the impact parameter is such that the following condition occurs
\begin{eqnarray}	
\hat{b}>(1-k \hat{d})
\label{IntervalS}
\end{eqnarray}
for some specific index $k$, all terms of (\ref{HalfDeflectionS_M}) with indexes $i>k$ vanish and only external rings with indexes $i<k$ contribute to the scattering angle. Overall, Fig.4 shows the deflection angles for positive and negative particles. The deflection functions are continuous functions of impact parameter for a certain interval of energies $E > \frac{R U_0}{d}$ or a certain interval of ring's radius $R < \frac{E d}{U_o}$, which illustrated in Fig.4(A),(C). The opposite case, 
\begin{eqnarray}	
R > R_{c}= \frac{2d}{\|\phi\|},
\label{CriticalRadius}
\end{eqnarray}
which we discussed in  Sec. \ref{sec:03} is also plotted in Fig.4(B),(D). 
In this case, the deflection functions are discontinuous, i.e. the incident coherent beam  is divided  by crystal's period into two separate parts. We can call them ``refracted''  and ``reflected'' parts, because ``refracted'' particles are mostly deflected to the convex side of ring and ``reflected'' particles deflected  to the opposite  side of the bend,  although it does not correspond to what is called ``refracted'' or ``reflected'' in optics. These parts are separated by a gap depended on potential shape. For our ring's model, the gap can be estimated from Eq.(\ref{HalfDeflectionS_M}). ``Reflected'' and ``refracted'' parts have  a distribution of angles rather than deflection at a single angle. 
In fact, the formula for distribution  (\ref{HalfDeflectionS_M}) allows to estimate the average angle for ``reflected'' and ``refracted'' fractions. For positive particles, this is an average integral over the interval $(1-i\hat{d})\sqrt{1-\phi}<\hat{b}<(1-i\hat{a})\sqrt{1-\phi}$. Because the deflection function is almost periodic, we can take second non distorted interval (Fig. 4(B)) with $i=1$:
\begin{eqnarray}	
\bar{\alpha}_{+} = \frac{1}{(\hat{d}-\hat{a})\sqrt{1-\phi}} \int^{(1-\hat{a})\sqrt{1-\phi}}_{(1-\hat{d})\sqrt{1-\phi}} 
(\frac{\hat{b}(\sqrt{1-\hat{b}^2}-\sqrt{1-\phi-\hat{b}^2})}{{\sqrt{1-\phi}}}-\nonumber\\
\frac{\hat{b}_{a}(\sqrt{1-\hat{b}^2_{a}}-\sqrt{1-\phi-\hat{b}^2_{a}})}{{\sqrt{1-\phi}}}+\nonumber\\ \frac{\hat{b}_{1}(\sqrt{1-\hat{b}_{1}^2})}{{\sqrt{1-\phi}}}) d\hat{b},
\label{AverageAngleP}
\end{eqnarray}
where $\hat{b}_{a}$, $\hat{b}_{1}$ are defined in (\ref{Normalize1}). This integral can be estimated, but the result is too cumbersome to reproduce it 
here. However, if we disregard the small terms  $\hat{a}^2$,$\hat{d}^2$,$\phi^2$,$\hat{a}\hat{d}$,$\phi\hat{d}$,$\phi\hat{a}$,etc., the final expression for average reflection angle $\bar{\alpha}_{+}$ will be:
\begin{eqnarray}	
\bar{\alpha}_{+} = \frac{1}{3(\hat{d}-\hat{a})}(2\phi^{3/2}-2\sqrt{2}\hat{d}^{3/2}+2\sqrt{2}\hat{a}^{3/2}+(2\hat{d}+\phi)^{3/2}-\nonumber\\
(2\hat{a}+\phi)^{3/2}+(2\hat{d}-2\hat{a})^{3/2}-(2\hat{d}-2\hat{a}+\phi)^{3/2}-(2\hat{a}-2\hat{d}+\phi)^{3/2}).
\label{AverageAnglePReflectF}
\end{eqnarray}
The experiment  for $70GeV$ protons in Si $<111>$ gives the average reflection angle, $39.5\mp2.0 \mu rad$  \cite{ivanov_2006}. 
Using the data $\phi=\theta_{c}^{2}=0.58*10^{-9}$, $a=0.78 \AA$, $d=3.13 \AA$, the formula (\ref{AverageAnglePReflectF}) gives $\chi_{+}=2*\bar{\alpha}_{+}=37.6 \mu rad$. This result is lower than experimental, but it stays inside the interval of experimental accuracy. 

Now, we estimate the ``refracted'' part of the angle distribution for positive particles. The integral for average angle will be
\begin{eqnarray}	
\bar{\alpha}_{-} = \frac{1}{\hat{a}\sqrt{1-\phi}} \int^{(1-\hat{d})\sqrt{1-\phi}}_{(1-\hat{a}-\hat{d})\sqrt{1-\phi}} 
(\frac{\hat{b}(\sqrt{1-\hat{b}^2}-\sqrt{1-\phi-\hat{b}^2})}{{\sqrt{1-\phi}}}-\nonumber\\
\frac{\hat{b}_{a}(\sqrt{1-\hat{b}^2_{a}}-\sqrt{1-\phi-\hat{b}^2_{a}})}{{\sqrt{1-\phi}}}+\nonumber\\ \frac{\hat{b}_{1}(\sqrt{1-\hat{b}_{1}^2}-\sqrt{1-\phi-\hat{b}_{1}^2})}{{\sqrt{1-\phi}}}-
\frac{\hat{b}_{1a}(\sqrt{1-\hat{b}_{1a}^2})}{{\sqrt{1-\phi}}}) d\hat{b}.
\label{AverageAnglePRefract}
\end{eqnarray}
The estimation of this integral is also straightforward, but due to cumbersome result  
we write down only its infinitesimal approximation of the first order:
\begin{eqnarray}	
\bar{\alpha}_{-} = \frac{1}{3\hat{a}}(-2\phi^{3/2}-2\sqrt{2}\hat{a}^{3/2}+4\sqrt{2}\hat{d}^{3/2}+(2\hat{a}+\phi)^{3/2}+\nonumber\\(-2\hat{a}+\phi)^{3/2}-(2\hat{d}+2\hat{a})^{3/2}-(2\hat{d}-2\hat{a})^{3/2}-\nonumber\\
(2\hat{d}+\phi)^{3/2}+(2\hat{a}+2\hat{d}+\phi)^{3/2}+(-2\hat{a}+2\hat{d}+\phi)^{3/2}).
\label{AverageAnglePRefractF}
\end{eqnarray}
The same experimental parameters  for $70GeV$ protons in Si $<111>$ gives the negative average ``refraction'' angle $\chi_{-}=2*\bar{\alpha}_{-}=-10.8 \mu rad$.  We may guess here that the fraction of particles on the right side of point B (see the emulsion's picture in \cite{ivanov_2006}) may be treated as a volume ``refraction'' effect. It is separated from the channeling fraction and has a smaller average angle than the ``reflection'' angle. However, a proof (or disproof) of this is desirable.

Now let us turn to the scattering of negative particles. The deflection function is described by the same equation (\ref{HalfDeflectionS_M}) with $\phi<0$. As a function of impact parameter $\hat{b}$, it will cause the mirror reflection relatively the axis $O \hat{b}$ and shift by the value $\sqrt{1-\phi}$ in the direction of $\hat{b}$ (see Fig.4(C),(D)). The corresponding fractions of scattering change its sign. In particular, the average ``reflected'' and ``refracted'' angles are given by the following equations:
\begin{eqnarray}	
\bar{\alpha}_{+} = \frac{1}{3\hat{a}}(2(-\phi)^{3/2}+2\sqrt{2}\hat{a}^{3/2}-4\sqrt{2}\hat{d}^{3/2}-(2\hat{a}-\phi)^{3/2}-\nonumber\\(-2\hat{a}-\phi)^{3/2}+(2\hat{d}+2\hat{a})^{3/2}+(2\hat{d}-2\hat{a})^{3/2}+\nonumber\\
(2\hat{d}-\phi)^{3/2}-(2\hat{a}+2\hat{d}-\phi)^{3/2}-(-2\hat{a}+2\hat{d}-\phi)^{3/2}).
\label{AverageAngleNReflectF}
\end{eqnarray}

\begin{eqnarray}	
\bar{\alpha}_{-} = \frac{1}{3(\hat{d}-\hat{a})}(-2(-\phi)^{3/2}+2\sqrt{2}\hat{d}^{3/2}-2\sqrt{2}\hat{a}^{3/2}-(2\hat{d}-\phi)^{3/2}+\nonumber\\
(2\hat{a}-\phi)^{3/2}-(2\hat{d}-2\hat{a})^{3/2}+(2\hat{d}-2\hat{a}-\phi)^{3/2}+(2\hat{a}-2\hat{d}-\phi)^{3/2}).
\label{AverageAngleNRefractF}
\end{eqnarray}

For negative particles, the ``refraction'' part is dominative and average angle of ``refraction'' (\ref{AverageAngleNRefractF}) for  $70GeV$ antiprotons in Si $<111>$ is $\chi_{-}=2*\bar{\alpha}_{-}=-37.6 \mu rad$. The ``reflective'' part of the antiproton  beam (\ref{AverageAngleNReflectF}) will have small but positive average angle $\chi_{+}=2*\bar{\alpha}_{+}=10.8 \mu rad$.

\section{\label{sec:06} Conclusion}

We have found the deflection function (Eqs. (\ref{HalfDeflectionS}), (\ref{HalfDeflectionGen_i}), (\ref{HalfDeflectionS_M})) for relativistic scattering by the system of periodic rectangular rings. Analysis of this function shows that an incident collimated beam is divided  by each crystal's period into two fractions: ``reflected'' and ``refracted''. These fractions are separated by a gap depended on potential shape, range of energy or radius of the bend.  The particles in the ``refracted'' fraction are mostly deflected to the side of the ring's bend. We receive analytical expression for average ``reflected'' angles (Eqs.(\ref{AverageAnglePReflectF}),(\ref{AverageAngleNReflectF})) which seems in satisfactory agreement with experiment.  We also receive expression for average ``refracted'' angles (Eqs.(\ref{AverageAnglePRefractF}),(\ref{AverageAngleNRefractF})).
The ``refracted'' fraction is smaller than ``reflected'' for  positive particles, but it might be present in  scattering data. This fraction should dominate in the scattering of the negative particles.

Author is grateful to Yuri M. Ivanov for the technical report \cite{ivanov_2005}.

\bibliography{../../Focusing_and_Channeling_in_Crystals/chan02}

\end{document}